\documentclass[sn-aps,iicol]{sn-jnl}
\usepackage[english]{babel}
\usepackage{lmodern}
\usepackage{microtype}
\usepackage{booktabs}
\usepackage{bm}
\usepackage{listings}
\usepackage{physics}
\usepackage{subfigure}
\usepackage{fancyhdr}
\usepackage{xcolor}
\usepackage{amssymb}
\usepackage{natbib}
\usepackage{orcidlink}

\newcommand{\pobs}{p_\mathrm{obs}}
\newcommand{\pint}{p_\mathrm{int}}
\newcommand{\pdet}{p_\mathrm{det}}
\renewcommand{\det}{\mathbb{D}}
\newcommand{\dphi}{\delta\phi}
\newcommand{\dhatphi}{\delta\hat{\phi}}
\newcommand{\detstat}{\mathbb{S}}

\begin{document}
\title{Accounting for selection biases in population analyses: equivalence of the in-likelihood and post-processing approaches}
\author*[1,2]{\fnm{Stefano} \sur{Rinaldi} \orcidlink{0000-0001-5799-4155}\,}\email{stefano.rinaldi@uni-heidelberg.de}
\affil[1]{\orgname{Institut für Theoretische Astrophysik, ZAH, Universität Heidelberg}, \orgaddress{\street{Albert-Ueberle-Stra{\ss}e 2}, \city{Heidelberg}, \postcode{D-69120}, \country{Germany}}}
\affil[2]{\orgname{Dipartimento di Fisica e Astronomia ``G. Galilei'', Università di Padova}, \orgaddress{\street{Via Marzolo 8}, \city{Padova}, \postcode{I-35131}, \country{Italy}}}

\date{\today}

\abstract{
In this paper, I show the equivalence, under appropriate assumptions, of two alternative methods to account for the presence of selection biases (also called selection effects) in population studies: one is to include the selection effects in the likelihood directly; the other follows the procedure of first inferring the observed distribution and then removing selection effects \emph{a posteriori}. Moreover, I investigate a potential bias allegedly induced by the latter approach: I show that this procedure, if applied under the appropriate assumptions, does not produce the aforementioned bias.
}

\maketitle

\section{Introduction}
Characterising the population-level properties of a catalogue of astronomical observations is crucial to understand the physical phenomena at play and to constrain the physics that governs our Universe. Unfortunately, the limited sensitivity of our instruments -- either electromagnetic telescopes or gravitational-wave (GW) interferometers -- introduces a bias in the observed population, which shows a preference for the easier-to-detect sources. This bias, discussed for the first time in Ref.~\citep{malmquist:1922}, goes under the name of \emph{selection bias} or, equivalently, \emph{selection effects}. Working within a Bayesian framework allows, on top of a robust treatment of measurement uncertainties, for the inclusion of selection effects in the inference scheme, effectively shielding the results from the selection bias. 

In what follows, I will restrict myself to the context of GW astronomy. Several works (e.g., \citealp{loredo:2004,mandel:2019}) elaborate on the subject, introducing frameworks to estimate the parameters of the intrinsic distributions in presence of selection effects under the assumption of having properly characterised the detector response, or in other words assuming the knowledge of the so-called \emph{selection function}.
The detector performance is studied via a set of injections produced by the LIGO-Virgo-KAGRA (LVK) collaboration \citep{sensitivityestimate:2021}, where a series of synthetic signals are injected in real detector noise and then flagged according to whether the search pipelines are able to detect each specific event or not. This injection set can be used either in the inference directly \citep{tiwari:2018,farr:2019,essick:2021} or to calibrate a (semi-)analytical approximant for the selection function \citep{wysocki:2019,veske:2021,lorenzo:2024}.

Recently, the non-parametric approach to population inference started to gain popularity, due to its flexibility in modelling arbitrary probability density without being committal to any specific functional form \citep{tiwari:2021:vamana,toubiana:2023,callister:2023,edelman:2023,sadiq:2023}. The idea of a non-parametric inference is to aim at reconstructing the shape of the distribution that most likely generated the available data, \emph{letting the data speak for themselves} in a certain sense. Overall, this is the most phenomenological approach possible: a direct interpretation of the (potentially infinite) parameters of a non-parametric model in terms of physical processes is not possible, and the inferred distribution can only be used as a guide for further developing parametric models.
My personal contribution to the field comes with \emph{a Hierarchy of Dirichlet Process Gaussian Mixture Models}, or (H)DPGMM for short \citep{rinaldi:2022:hdpgmm}, a hierarchical model that builds on the properties of the Dirichlet process Gaussian mixture model (DPGMM). We applied this model to the problem of reconstructing the intrinsic distribution of the primary mass, mass ratio, and redshift of the binary black hole (BBH) population using the GW events detected during the three LVK observing run concluded to date \citep{rinaldi:2023:m1qz}. 

Due to the specific formulation of the Dirichlet process (DP), this method is primarily designed to reconstruct the so-called \emph{observed distribution}. The selection effects are then removed in post-processing\footnote{In this paper, I will call \emph{intrinsic distribution} the one from which the observed events are drawn before accounting for selection effects and \emph{observed distribution} the one after the application of selection effects.}:
\begin{multline}\label{eq:postprocess}
\pobs(\theta|\Lambda)\propto \pint(\theta|\Lambda)\pdet(\theta) \\ \Rightarrow \pint(\theta|\Lambda) \propto \frac{\pobs(\theta|\Lambda)}{\pdet(\theta)}\,.
\end{multline}
Here I denote with $\pint$ the intrinsic distribution, with $\pobs$ the observed distribution and with $\pdet$ the selection function. $\theta$ denotes the BBH parameters (e.g., primary mass and redshift) and $\Lambda$ are the parameters of the intrinsic distribution.
Ref.~\citep{essick:2023} claims that this way of accounting for selection effects is not correct and might lead to a biased inference, especially when applied to a large number of events.

In this paper, I will show that, under appropriate assumptions, this approach is equivalent to the one presented in Ref.~\citep{mandel:2019} and that a proper treatment of the observed distribution results in an unbiased inference of the intrinsic distribution. Section~\ref{sec:derivation} describes the role of the observed distribution parameters and Section~\ref{sec:toymodel} presents the application of this scheme to a simple toy problem.

\section{Statistical framework}\label{sec:derivation}
The observed distribution can be thought as the distribution of the parameters $\theta$ conditioned on the fact that the observations are detected, denoted with $\det$. Making use of the Bayes' theorem, this distribution reads
\begin{equation}\label{eq:bayestheorem}
    p(\theta|\det) = \frac{p(\det|\theta)p(\theta)}{p(\det)}\,.
\end{equation}
These terms can be identified as follows:
\begin{itemize}
    \item $p(\theta|\det)$ is the observed distribution $\pobs(\theta|\Lambda)$;
    \item $p(\theta)$ is the intrinsic distribution $\pint(\theta|\Lambda)$;
    \item $p(\det|\theta)$ is the detection probability of an event with parameters $\theta$, the selection function $\pdet(\theta)$. In principle, it depends on its own set of parameters: in what follows, I will assume a perfect knowledge of both its functional form and parameters, omitting to indicate them in the derivation for the sake of brevity;
    \item $p(\det)$ is the detection probability marginalised over all the possible parameters $\theta$. This term will be absorbed in an overall normalisation constant.
\end{itemize}
From Eq.~\eqref{eq:bayestheorem} follows Eq.~\eqref{eq:postprocess}:
\begin{equation}\label{eq:pastropdet}
    \pobs(\theta|\Lambda) \propto \pint(\theta|\Lambda)\pdet(\theta)\,.
\end{equation}
In general it is not possible to write a non-trivial\footnote{Here \emph{non-trivial} means ``different from the explicit product $\pint(\theta|\Lambda)\pdet(\theta)$''.} closed form for $\pobs(\theta|\Lambda)$, thus one has to infer the parameters of the intrinsic distribution $\Lambda$ accounting for selection effects as described in the papers cited in the previous Section \citep{mandel:2019,essick:2023} and include the selection function in the likelihood derivation (hence the name \emph{in-likelihood approach}). There are two situations in which it is possible to infer the observed distribution directly: the first is when the product of the intrinsic distribution and the selection function can be expressed in a closed form, usually when the two functions are conjugated, and the second is when an effective approximant for $\pobs(\theta)$ is available. In both cases, the inferred parameters are not the parameters of the intrinsic distribution $\Lambda$ but a second set of parameters, which I will denote with $\Theta$, that are used to describe the observed distribution, $\pobs(\theta|\Theta)$. This distinction is not a mere matter of naming conventions but, as I will show in the following, the difference between $\Lambda$ and $\Theta$ is crucial.

\subsection{Closed form for \texorpdfstring{$\mathbf{\pobs}$}{}}\label{sec:closedform}
If the product in Eq.~\eqref{eq:pastropdet} can be expressed in a closed form with its own parameters it is possible to exploit this fact to infer the observed distribution parameters $\Theta$ and then map $\Theta$ in the intrinsic parameters $\Lambda$. A simple case where this happens is when the intrinsic distribution and the selection function are conjugated.
In the context of Bayesian probability theory, a likelihood function is said to be conjugated to the prior distribution if the resulting posterior is in the same functional family of the prior with some updated parameters, hence retains the same functional form of the intrinsic distribution. If this is the case, it is possible to write a closed form for the observed distribution and to link the parameters of the latter, $\Theta$, to the intrinsic parameters $\Lambda$:
\begin{multline}
    \pint(\theta|\Lambda)\pdet(\theta) = F(\theta|\Lambda)\pdet(\theta) \\ = F\qty(\theta|\Theta(\Lambda))= \pobs\qty(\theta|\Lambda)\,.
\end{multline}
Here $F(\theta)$ is the conjugated distribution and $\Theta(\Lambda)$ denotes the functional relation between $\Theta$ and $\Lambda$. The conjugation between the intrinsic distribution and the selection function opens for the possibility of inferring the parameters $\Theta$ using the available observations and accounting for selection effects \emph{a posteriori}, inverting the relation between $\Theta$ and $\Lambda$: the relation $\Lambda(\Theta) = \Theta^{-1}(\Lambda)$ properly accounts for selection effects, analytically ``dividing for the selection function''. Conversely, if one does not properly re-map the inferred parameters into the intrinsic distribution parameters properly incurs in a bias, since the inferred parameters are not meant to describe the intrinsic distribution. In Section~\ref{sec:toymodel}, I show that accounting for the relation between $\Theta$ and $\Lambda$ prevents the occurrence of such bias with an analogous toy problem.

The bias presented in Section~4.1 of Ref.~\citep{essick:2023} is a logic consequence, in a certain sense, of the discussion above: if the selection function is not conjugated to the intrinsic distribution, there is no reason to assume that the observed distribution $\pobs(\theta|\Theta)$ will have the same functional form and parameters of $\pint(\theta|\Lambda)$. The two distributions will be different, and a closed form for the observed distribution might not even exist. In this situation, trying to fit the observations with the functional form of the intrinsic distribution and then dividing by the selection function will inevitably introduce a bias simply because the employed distribution is not a good descriptor for the data. 

There are two ways to circumvent this bias in absence of a closed form for the observed distribution: the first is to use the framework described in Ref.~\citep{essick:2023}, which closely follows Ref.~\citep{mandel:2019} and includes the selection effects in the likelihood function; the second makes use of an approximant for the observed distribution, and is described in the following Section.

\subsection{Approximant for \texorpdfstring{$\mathbf{\pobs}$}{}}\label{sec:approximant}
In the absence of a closed form for $\pobs(\theta)$, one can make use of approximants informed with the available observations. This can be done making use of a variety of approaches, such as machine learning or non-parametric methods. As long as the condition
\begin{equation}\label{eq:approximant}
    \pobs(\theta|\Lambda) \simeq A(\theta|\Theta)\,,
\end{equation}
where $A(\theta|\Theta)$ denotes the approximant function with parameters $\Theta$, is met, it is possible to build the approximant $A(\theta|\Theta)$ using the available data and then remove the selection effects in post-processing as in Eq.~\eqref{eq:postprocess}:
\begin{equation}
    \pint(\theta|\Lambda) \propto \frac{\pobs(\theta|\Lambda)}{\pdet(\theta)} \simeq \frac{A(\theta|\Theta)}{\pdet(\theta)}\,.
\end{equation}
It goes without saying that in this case the intrinsic parameters $\Lambda$ have no relation with the approximant parameters $\Theta$, and the intrinsic distribution is represented by the ratio between the reconstructed observed distribution and the selection function.

The approach taken in Ref.~\citep{rinaldi:2022:hdpgmm}, makes use of this second prescription and approximates the probability density conditioned on the detection, $p(\theta|\det)$: the DPGMM ensures that the condition expressed in Eq.~\eqref{eq:approximant} is met, as demonstrated in Ref.~\citep{nguyen:2020}. It is however possible to include the selection effects directly in a DPGMM-based model such as (H)DPGMM following the approach of Ref.~\citep{mandel:2019}. More details can be found in Appendix~\ref{app:dpgmm-sf}.

\subsection{Parameter estimation likelihood}\label{sec:parest}
Estimating the parameters of a given model -- either $\Lambda$ describing the intrinsic distribution or $\Theta$ for the observed one -- requires the derivation of the likelihood function, or the probability of observing the available data given $\Lambda$ or $\Theta$. For this problem, the likelihood function is slightly different depending on whether one is considering the intrinsic or the observed distribution: this difference stems from the fact that the observed distribution $\pobs(\theta|\Theta,\det)$ is already conditioned on the fact that the observations are detectable (and detected), whereas the intrinsic distribution $\pint(\theta|\Lambda)$ is not. In other words, it depends on which side of the equation
\begin{equation}
\pint(\theta|\Lambda) = \frac{\pobs(\theta|\Theta,\det)}{\pdet(\theta)}p(\det)
\end{equation}
the inference is based upon\footnote{The inconsistency between the parameters on the two sides of the equation is deliberate and meant to highlight that the two distributions, intrinsic and observed, can have different functional forms and parameters.}. Assuming a model for the intrinsic distribution means that the selection effects are not included in the model and they have to be included in the likelihood, whereas a functional form (or, equivalently, an approximant) for the observed distribution is already conditioned on the fact that only a fraction of the objects are observable.

The likelihood for the in-likelihood approach $p(\mathbf{d}|\Lambda,\det)$ is derived in Refs.~\citep{mandel:2019,essick:2023}, and it must take into account the fact that the data $\mathbf{d} = \{d_1,\ldots,d_N\}$ are filtered by the selection function before the observation. Here I briefly summarise their derivation using the formalism of this paper. Assuming that each observation $d_i$ is independent and identically distributed (i.i.d.),
\begin{equation}\label{eq:intrinsic_likelihood}
p(\mathbf{d}|\Lambda,\det) = \prod_i p(d_i|\Lambda,\det)\,,
\end{equation}
and that its likelihood depends only on the parameters of the individual observation $\theta_i$ -- the noise distribution $p(d_i|\theta_i)$ -- the likelihood is
\begin{equation}
p(\mathbf{d}|\Lambda,\det) = \prod_i\int p(d_i|\theta_i,\det)p(\theta_i|\Lambda,\det)\dd\theta_i\,.
\end{equation}
The integrand can be rewritten as
\begin{multline}
\frac{p(\det|d_i,\theta_i)p(d_i|\theta_i)}{p(\det|\theta_i)}\frac{p(\det|\theta_i)p(\theta_i|\Lambda)}{p(\det|\Lambda)} \\= \frac{p(d_i|\theta_i)p(\theta_i|\Lambda)}{p(\det|\Lambda)}\,,
\end{multline}
where I dropped $\Lambda$ from both $p(d_i|\theta_i,\Lambda)$ and $p(\det|\theta_i,\Lambda)$ and I used the fact that $p(\det|d_i,\theta_i) = 1$ by definition, since it is conditioned on some observed data $d_i$. This is based on the assumption that the detection is a threshold process based on some deterministic detection statistic obtained from the data $d_i$.
The term $p(\det|\Lambda)$ is the $\alpha(\Lambda)$ factor defined in Eq.~(6) of Ref.~\citep{mandel:2019}:
\begin{equation}\label{eq:lambda_MFG}
p(\det|\Lambda) = \int p(\det|\theta)p(\theta|\Lambda)\dd\theta \equiv \alpha(\Lambda)\,.
\end{equation}
The likelihood in Eq.~\eqref{eq:intrinsic_likelihood} then becomes
\begin{equation}\label{eq:MFG_likelihood}
p(\mathbf{d}|\Lambda,\det) = \qty(\frac{1}{\alpha(\Lambda)})^N \prod_i\int p(d_i|\theta_i)p(\theta_i|\Lambda) \dd \theta_i\,.
\end{equation}

For the observed distribution parameters $\Theta$, on the other hand, the condition on the detection $\det$ is already included in the assumptions made about the distribution: the observed data are distributed according to the already filtered $p(\theta|\Theta,\det)$. Therefore, the likelihood $p(\mathbf{d}|\Theta)$ is not conditioned on $\det$ because all the realisations from this distribution are, by definition, detectable. Under the assumption of i.i.d. observations, the likelihood simply reads
\begin{equation}\label{eq:observed_likelihood}
p(\mathbf{d}|\Theta) = \prod_i \int p(d_i|\theta_i)p(\theta_i|\Theta) \dd\theta\,.
\end{equation}
What is not encoded in the post-processing approach, however, is the potential presence of a correlation between the detection statistic and the measured parameters of the detected data\footnote{In other words, the distribution of the percentile at which the parameters $\theta_i$ are found in the posterior distribution $p(\theta_i|d_i)$ has to be uniform considering the selected observations only. If this is not the case, this assumption is not met.}. Since the correction for selection effects is accounted for in the observed distribution rather than included for each individual event directly, it is not possible to include this effect in the post-processing likelihood. If this assumption is not met, the post-processing approach does not apply and one has to rely on the in-likelihood approach. This requirement, however, does not stem from the fact that the detection $\det$ is a function of the parameters $\theta$ alone: the detection probability $\pdet(\theta)$ includes the fact that a particular set of parameters $\theta_i$ may or may not be detectable depending on the noise $n_i$ -- hence on the data $d_i$.

\section{Toy models}\label{sec:toymodel}
In this Section, I will apply the two prescriptions presented in Sections~\ref{sec:closedform} and~\ref{sec:approximant} to the toy problems presented in Section~4.2 and Appendixes~B.1 and B.2 of Ref.~\citep{essick:2023}, of which I will follow the notation.

\subsection{Gaussian distribution}
This example presents a model for parametrised deviations from General Relativity with a single deviation parameter, $\dphi$. The equivalent of the intrinsic distribution for the deviation parameter is modelled as a Gaussian distribution with mean $\mu_\Lambda$ and standard deviation $\sigma_\Lambda$:
\begin{equation}
    \pint(\dphi|\Lambda) = \mathcal{N}(\dphi|\mu_\Lambda,\sigma_\Lambda)\,,
\end{equation}
where $\Lambda = \{\mu_\Lambda,\sigma_\Lambda\}$. The probability of detecting an event is given by the selection function, which is assumed to be Gaussian with known parameters $\mu_D$ and $\sigma_D$:
\begin{equation}
    \pdet(\dphi) = \mathcal{N}(\dphi|\mu_D,\sigma_D)\,.
\end{equation}
An event with deviation parameter $\dphi_i$,
\begin{equation}
\dphi_i\sim\pint(\dphi|\Lambda)\,,
\end{equation}
is included in our simulated catalogue with probability $\pdet(\dphi_i)$. This is the only departure point from the model presented in Ref.~\citep{essick:2023}: they model their detection probability as a function of the observed value $\dhatphi_i$, whereas in this framework the selection function depends on $\dphi_i$, consistently with a deterministic threshold-based selection process such as the GW detection. This point is discussed in Appendix~\ref{app:selfunc}. Without this difference, the assumptions required by the post-processing approach are not met.
The observed distribution is the product of these two functions:
\begin{multline}
    \pobs(\dphi|\Lambda) \propto \pint(\dphi|\Lambda)\pdet(\dphi) \\ = \mathcal{N}(\dphi|\mu_\Lambda,\sigma_\Lambda)\mathcal{N}(\dphi|\mu_D,\sigma_D)\\ \propto \mathcal{N}(\dphi|\mu_\mathrm{obs},\sigma_\mathrm{obs})\,,
\end{multline}
where
\begin{equation}\label{eq:thetaoflambda_mu}
    \mu_\mathrm{obs} = \frac{\sigma_D^2\mu_\Lambda + \sigma_\Lambda^2\mu_D}{\sigma_D^2+\sigma_\Lambda^2}\,,
\end{equation}
and
\begin{equation}\label{eq:thetaoflambda_sigma}
    \sigma_\mathrm{obs} =\qty(\frac{1}{\sigma_D^2}+\frac{1}{\sigma_\Lambda^2})^{-1/2}\,.
\end{equation}
In this example the two distributions are conjugated, so it is possible to write a closed form for $\pobs$ and a relation between $\Lambda$ and $\Theta = \{\mu_\mathrm{obs}, \sigma_\mathrm{obs}\}$.
For each event included in the catalogue I then draw a measured value $\dhatphi_i$ from a Gaussian distribution centred on $\dphi_i$ with known standard deviation $\sigma_i$:
\begin{equation}
    \dhatphi_i \sim \mathcal{N}(\dhatphi|\dphi_i,\sigma_i)\,.
\end{equation}
This procedure produces a set of $N$ measured deviation parameters $\Phi = \{\dhatphi_1,\ldots,\dhatphi_N\}$ along with their associated uncertainty $\Sigma = \{\sigma_1,\ldots,\sigma_N\}$. Moreover, I assume a perfect knowledge of the selection function parameters $\mu_D$ and $\sigma_D$.

The observed distribution parameters $\Theta = \{\mu_\mathrm{obs},\sigma_\mathrm{obs}\}$ follows:
\begin{multline}\label{eq:posterior}
    p(\mu_\mathrm{obs},\sigma_\mathrm{obs}|\Phi,\Sigma) \propto p(\Phi|\Sigma,\mu_\mathrm{obs},\sigma_\mathrm{obs})p(\mu_\mathrm{obs},\sigma_\mathrm{obs}) \\ = p(\mu_\mathrm{obs},\sigma_\mathrm{obs})\prod_i p(\dhatphi_i|\mu_\mathrm{obs},\sigma_\mathrm{obs})\,.
\end{multline}
The individual observation likelihood reads
\begin{multline}
    p(\dhatphi_i|\mu_\mathrm{obs},\sigma_\mathrm{obs}) \\= \int p(\dhatphi_i|\dphi_i,\sigma_i)p(\dphi_i|\mu_\mathrm{obs},\sigma_\mathrm{obs}) \dd \dphi_i \\= \mathcal{N}\qty(\dhatphi_i\bigg\rvert\mu_\mathrm{obs},\sqrt{\sigma_\mathrm{obs}^2+\sigma_i^2})\,,
\end{multline}
where in the last equality I carried out the integration analytically. In the following, I will make the simplifying assumption that $\sigma_i = \sigma_0$ for all the observations.
The prior distribution $p(\mu_\mathrm{obs},\sigma_\mathrm{obs})$ is assumed to be uniform.
The posterior distribution for $\mu_\mathrm{obs}$ and $\sigma_\mathrm{obs}$ can be explored making use of a stochastic sampler: in particular, I used \textsc{RayNest}\footnote{Publicly available at \url{https://github.com/wdpozzo/raynest}.}, an implementation of the nested sampling scheme \citep{skilling:2006}. The posterior samples for $\mu_\mathrm{obs}$ and $\sigma_\mathrm{obs}$ are then converted into posterior samples for $\mu_\Lambda$ and $\sigma_\Lambda$ inverting Eqs.~\eqref{eq:thetaoflambda_mu} and~\eqref{eq:thetaoflambda_sigma}.

\begin{table}
    \centering
    \caption{Parameters for the three different populations used in the toy model example. The first two, \emph{wide} and \emph{narrow}, follows the two models given in the caption of Figure~5 of Ref.~\citep{essick:2023}, whereas the third, \emph{equal}, is designed to avoid all the limits in which the bias is expected to be negligible (Appendix~B.1 of Ref.~\citep{essick:2023}).}
    \begin{tabular}{lccc}
        \toprule
         & Wide & Narrow & Equal  \\
        \midrule
        $\mu_\Lambda$ & -2 & -2 & -2 \\
        $\sigma_\Lambda$ & 3 & 0.6 & 1 \\
        $\mu_D$ & 0 & 0 & 0 \\
        $\sigma_D$ & 2 & 2 & 1 \\
        $\sigma_0$ & 1 & 1 & 1 \\
        \bottomrule
    \end{tabular}
    \label{tab:toymodelpars}
\end{table}

\begin{figure}
    \centering
    \includegraphics[width=0.8\columnwidth]{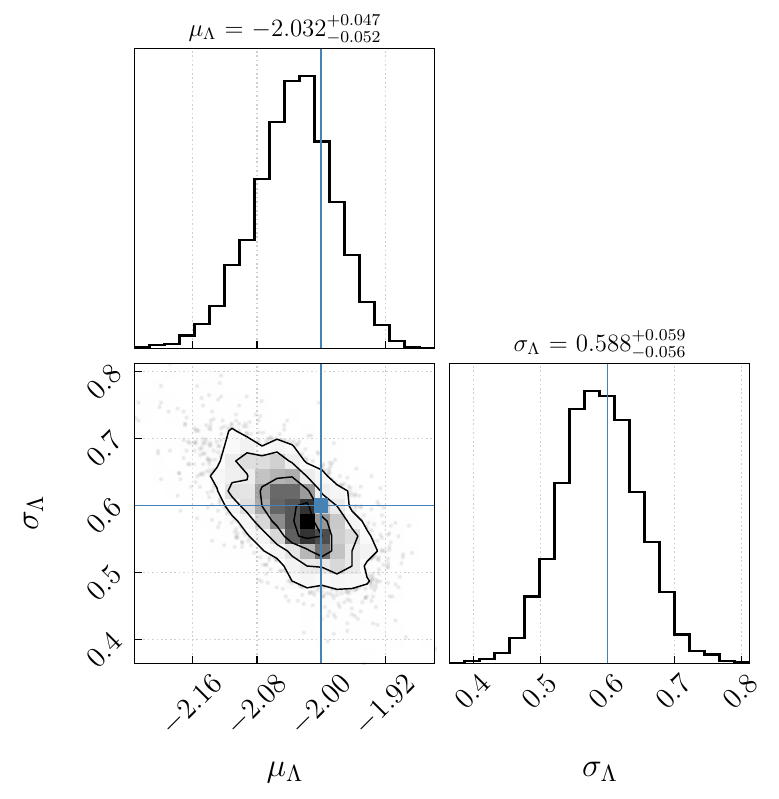}
    \caption{Posterior distribution for $\mu_\Lambda$ and $\sigma_\Lambda$ with $1000$ observations, \emph{narrow} population model. The blue lines mark the simulated values.}
    \label{fig:posterior_1e3}
\end{figure}

The first test I present here follows the \emph{narrow} population model presented in Section~4.2 of Ref.~\citep{essick:2023} (see their Figure~5 for the details). The parameters of this population model are given in Table~\ref{tab:toymodelpars}. I simulated 1000 observations and explored the posterior distribution for $\Theta = \{\mu_\mathrm{obs}, \sigma_\mathrm{obs}\}$ given in Eq.~\eqref{eq:posterior}. The samples are then transformed into samples for $\Lambda = \{\mu_\Lambda,\sigma_\Lambda\}$: the recovered posterior distribution is reported in Figure~\ref{fig:posterior_1e3}, and it is consistent with the simulated values.

\begin{figure*}
    \centering
    \includegraphics[width=1.8\columnwidth]{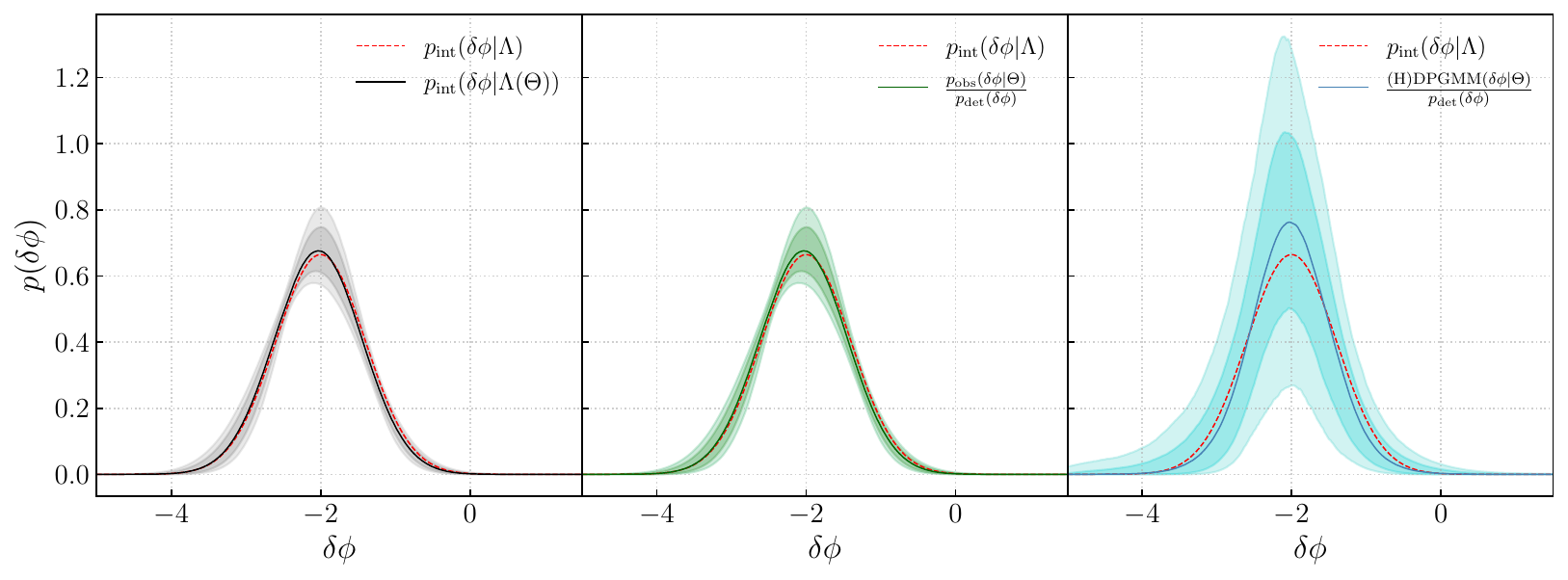}
    \caption{Recovered intrinsic distribution (\emph{narrow} model) with three different prescriptions compared with the simulated $\pint(\dphi|\Lambda)$. From left to right: remapping the observed parameters $\Theta$ into the intrinsic parameters $\Lambda$, dividing the observed distribution by the selection function and approximating the observed distribution with a non-parametric method, (H)DPGMM, before dividing by the selection function.}
    \label{fig:comparison}
\end{figure*}

Figure~\ref{fig:comparison} compares the recovered posterior distribution for the intrinsic distribution with three different recipes:
\begin{enumerate}
    \item Remapping $\Theta$ into $\Lambda$;
    \item Dividing the inferred observed distribution $\pobs(\dphi|\Theta)$ by the selection function $\pdet(\dphi)$;
    \item Reconstructing $\pobs(\dphi|\Lambda)$ via (H)DPGMM\footnote{The reconstruction is made with \textsc{figaro} \citep{rinaldi:2022:figaro}, publicly available at \url{https://github.com/sterinaldi/FIGARO}.} and dividing the approximant by the selection function.
\end{enumerate}
These three distributions are all consistent with the intrinsic distribution within the statistical uncertainty.

To ensure the statistical robustness of this method, I also run a PP-plot test for each of the three choices listed in Table~\ref{tab:toymodelpars}. For each parameter choice I repeated 100 times the exercise of simulating 1000 observed events, inferring the parameters $\Theta$, remapping $\Theta$ into $\Lambda$ and then computing the percentile at which the true values of $\Lambda$ are found.
If a bias is present in this approach, the percentile distribution of the true value should deviate from the uniform distribution. This does not happen even for the \emph{equal} model, where $\sigma_\Lambda = \sigma_D = \sigma_0$ and $\mu_\Lambda \neq \mu_D$: this choice, according to Appendix~B.1 of Ref.~\citep{essick:2023}, should exhibit the largest bias.

\begin{figure}
    \centering
    \includegraphics[width=0.8\columnwidth]{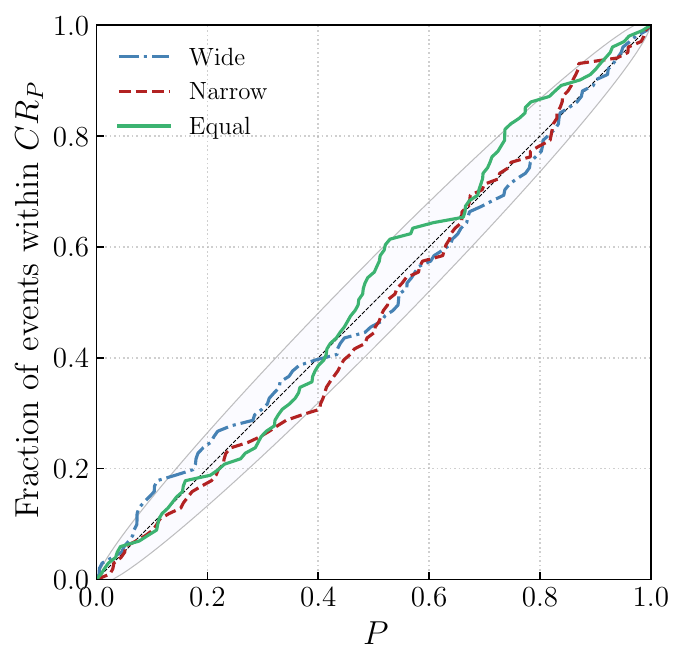}
    \caption{PP-plot for the three different sets of parameters for the toy model presented in Section~\ref{sec:toymodel}, comparing the expected and observed percentile (or credible region, CR) cumulative distributions. The shaded region represents the $90\%$ credible region for the cumulative distribution, estimated from a Beta distribution as in Ref.~\citep{cameron:2011}.}
    \label{fig:ppplot}
\end{figure}

\begin{figure}
    \centering
    \includegraphics[width=0.8\columnwidth]{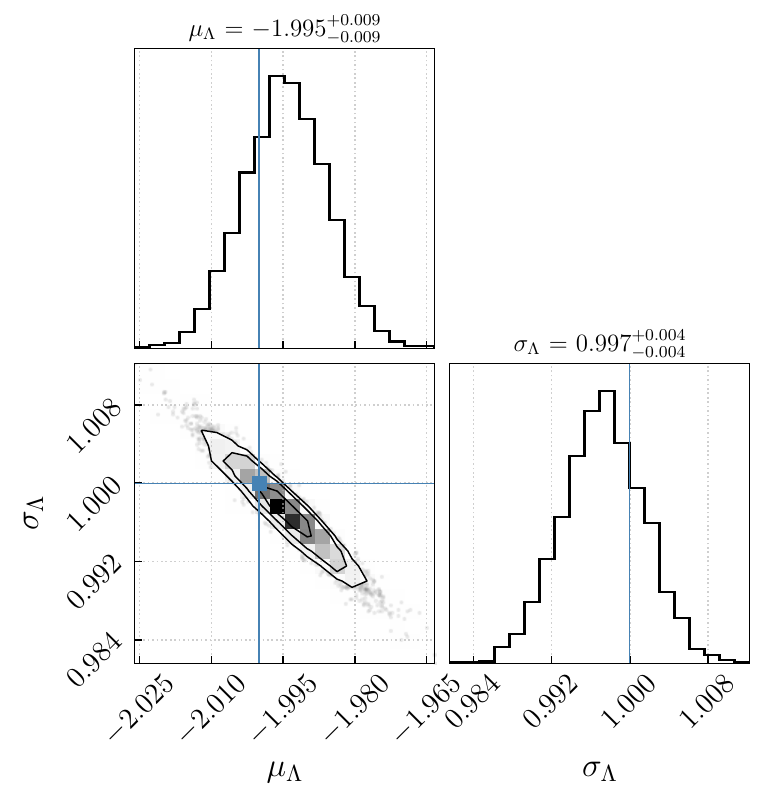}
    \caption{Posterior distribution for $\mu_\Lambda$ and $\sigma_\Lambda$ with $10^6$ observations, \emph{equal} population model. The blue lines mark the simulated values.}
    \label{fig:posterior_1e6}
\end{figure}

As a last test, I run the inference scheme using a catalogue of $10^6$ events from the \emph{equal} model. The larger number of available events reduces the uncertainty on the parameters, thus making the bias, if present, more evident. Figure~\ref{fig:posterior_1e6} reports the inferred posterior distribution obtained remapping $\Theta$ into $\Lambda$, and even in this case the posterior distribution is consistent with the simulated values.

\subsection{Truncated Gaussian}
In the example presented in the previous Section, the intrinsic and observed distribution share the same functional form. In general, however, this is not the case: an example of this can be found in Appendix~B.2 of Ref.~\citep{essick:2023}, where the selection function is assumed to be a step function:
\begin{equation}
\pdet(\delta\phi) = \begin{cases} 1 \qif \delta\phi > \delta\phi_\mathrm{th}\\ 0 \qif \delta\phi < \delta\phi_\mathrm{th} \end{cases}.
\end{equation}
Under the assumption of a Gaussian intrinsic distribution, the observed distribution reads
\begin{multline}
\pobs(\delta\phi|\Lambda) \propto \pint(\delta\phi|\Lambda)\pdet(\delta\phi)\\ \propto \mathcal{TN}(\delta\phi|\mu_\Lambda,\sigma_\Lambda,\delta\phi_\mathrm{th})\,,
\end{multline}
where $\mathcal{TN}(\delta\phi)$ denotes the truncated Gaussian distribution, defined as
\begin{equation}
\mathcal{TN}(\delta\phi) \propto \begin{cases} \mathcal{N}(\delta\phi|\mu_\Lambda,\sigma_\Lambda) \qif \delta\phi > \delta\phi_\mathrm{th}\\ 0 \qif \delta\phi < \delta\phi_\mathrm{th} \end{cases}.
\end{equation}

Here I show that making use of the prescription given in Section~\ref{sec:closedform}, hence modelling the observed data as a truncated Gaussian distribution, allows for a correct inference of the intrinsic distribution parameters even in presence of a completely vetoed region of the parameter space.

\begin{figure}
    \centering
    \includegraphics[width=\columnwidth]{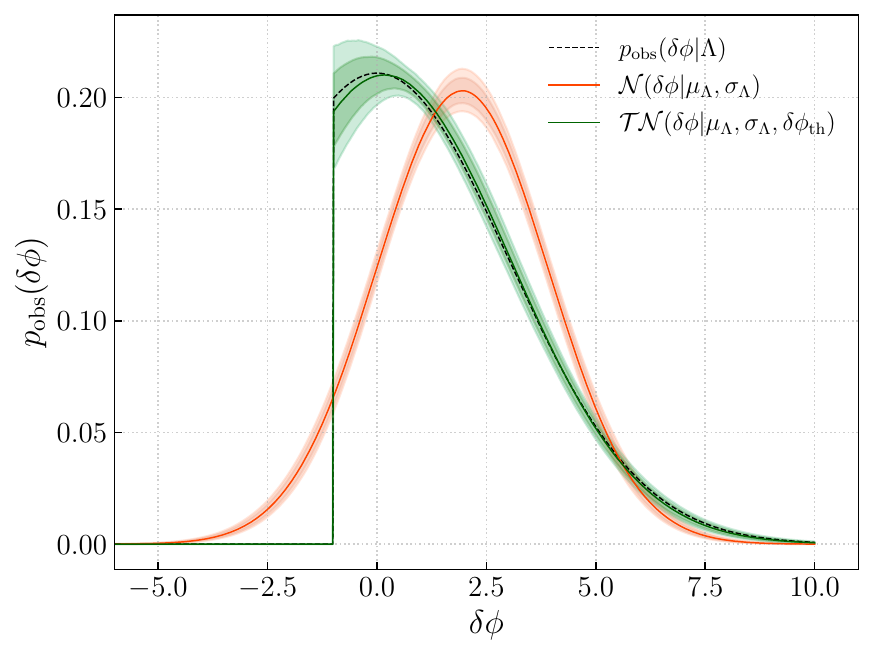}
    \caption{Recovered observed distribution for the step function selection effects example: the truncated Gaussian model (green) is consistent with the simulated $\pobs(\dphi|\Lambda)$ (dashed black), whereas the Gaussian model (red) is not.}
    \label{fig:truncated_gaussian}
\end{figure}

The test presented here simulates an intrinsic Gaussian distribution with $\mu_\Lambda = 0$, $\sigma_\Lambda = 3$, $\sigma_0 = 1$ and a selection function with $\delta\phi_\mathrm{th} = -1$. I simulated $10^3$ detections and analysed them both following the prescription given in Section~\ref{sec:closedform} and with the wrong approach of using the intrinsic distribution model directly. As before, I assume a perfect knowledge of the selection function. Figure~\ref{fig:truncated_gaussian} compares the inferred observed distribution with the two methods: in one case, the correct observed distribution is recovered, whereas making use of the wrong approach I find something analogous to the bias pointed out in Ref.~\citep{essick:2023}.

\section{Summary}
In this paper, I have presented the statistical framework that allows for the inclusion of selection effects \emph{a posteriori} with respect to the population inference using a reasonable observed distribution model. Under the assumption that the detection statistic does not correlate with the measured parameters directly but at most with their precision, this approach is capable of inferring the parameters of the underlying distribution.

The use of the method presented in this paper is not universal and it can be applied straightforwardly in just a handful of special cases, of which two are detailed in this paper: therefore, care must be taken while using this approach. Nonetheless, within the range of applicability of the assumptions made, this method can provide a useful alternative to the standard in-likelihood approach, where the selection effects are included in the inference scheme directly. 

\section*{Acknowledgements}
I thank Riccardo~Buscicchio, Walter~Del~Pozzo, Zoheyr~Doctor and Reed~Essick for useful discussions and comments.

\section*{Statements and declarations}
\subsection*{Funding}
I acknowledge financial support from the European Research Council for the ERC Consolidator grant DEMOBLACK, under contract no. 770017, and from the German Excellence Strategy via the Heidelberg Cluster of Excellence (EXC 2181 - 390900948) STRUCTURES.

\subsection*{Code availability}
\textsc{figaro} is publicly available at \url{https://github.com/sterinaldi/figaro} and \textsc{raynest} is publicly available at \url{https://github.com/wdpozzo/raynest}.
\subsection*{Data availability}
This manuscript has no associated data.

\subsection*{Competing interests}
The author certifies that he has no affiliations with or involvement in any organisation or entity with any financial interest or non-financial interest in the subject matter or materials discussed in this manuscript.

\appendix

\section{DPGMM inference with selection effects}\label{app:dpgmm-sf}
The parameters of a (potentially infinite and multivariate) Gaussian Mixture Model, or GMM, are a vector of weights $\mathbf{w} = \{w_1,w_2,\ldots\}$, a vector of means $\boldsymbol{\mu} = \{\mu_1,\mu_2,\ldots\}$ and a vector of covariance matrices $\boldsymbol\sigma =\{\sigma_1,\sigma_2,\ldots\}$. Methods for exploring this parameter spaces are available in literature, such as variational algorithms \citep{pedregosa:2011,delpozzo:2018} and Gibbs sampling schemes \citep{neal:2000,gorur:2010}. (H)DPGMM \citep{rinaldi:2022:hdpgmm} was derived with the collapsed Gibbs sampling approach in mind, thus making use of conditional posterior distributions. It accounts for the fact that individual observations are affected by noise including the posterior probability density in its likelihood, as in Eq.~\eqref{eq:observed_likelihood}, but it does not include the selection function in it, relying on the correction a posteriori: in this Appendix, I will outline how it is possible to account for the presence of selection biases during the inference itself making use of the results of Ref.~\citep{mandel:2019}, referring the interested reader to Ref.~\citep{rinaldi:2022:hdpgmm} for the derivation of the distributions without selection effects.

In a nutshell, the idea behind the collapsed Gibbs sampling exploration of the (H)DPGMM parameter space is that one should first group the available observations into clusters associated with the Gaussian mixture components: then, it is possible to sample from the probability density for $\mathbf{w},\boldsymbol\mu,\boldsymbol\sigma$ conditioned on the data $\mathbf{d}$ and the cluster association of each observation, denoted with $\mathbf{z} = \{z_1,z_2,\ldots,z_N\}$\footnote{The label $z_n=k$ means \emph{the $n-$th observation is associated with the $k-$th Gaussian component}.}.
\begin{equation}
\mathbf{w}_i,\boldsymbol\mu_i,\boldsymbol\sigma_i \sim p(\mathbf{w},\boldsymbol\mu,\boldsymbol\sigma|\mathbf{z},\mathbf{d})\,.
\end{equation}
Since the different mixture components are independent from each other, the probability for $\mu_i$ and $\sigma_i$ depends only on the observations associated with the $j-$th mixture component:
\begin{multline}
p(\mu_j,\sigma_j|\mathbf{z},\mathbf{d},\{i|z_i=j\}) \\ \propto p(\mu_j,\sigma_j)\prod_i \int \mathcal{N}(\theta_i|\mu_j,\sigma_j)p(\theta_i|d_i)\dd \theta_i\,,
\end{multline}
without including selection effects at this stage, and
\begin{multline}
p(\mu_j,\sigma_j|\mathbf{z},\mathbf{d},\{i|z_i=j\}, \det) \\ \propto p(\mu_j,\sigma_j) \prod_i \int \frac{\mathcal{N}(\theta_i|\mu_j,\sigma_j)p(\theta_i|d_i)}{\alpha(\mu_j,\sigma_j)}\dd \theta_i
\end{multline}
with the inclusion of selection effects as in Ref.~\citep{mandel:2019}. 

\begin{figure}
    \centering
    \includegraphics[width=0.8\columnwidth]{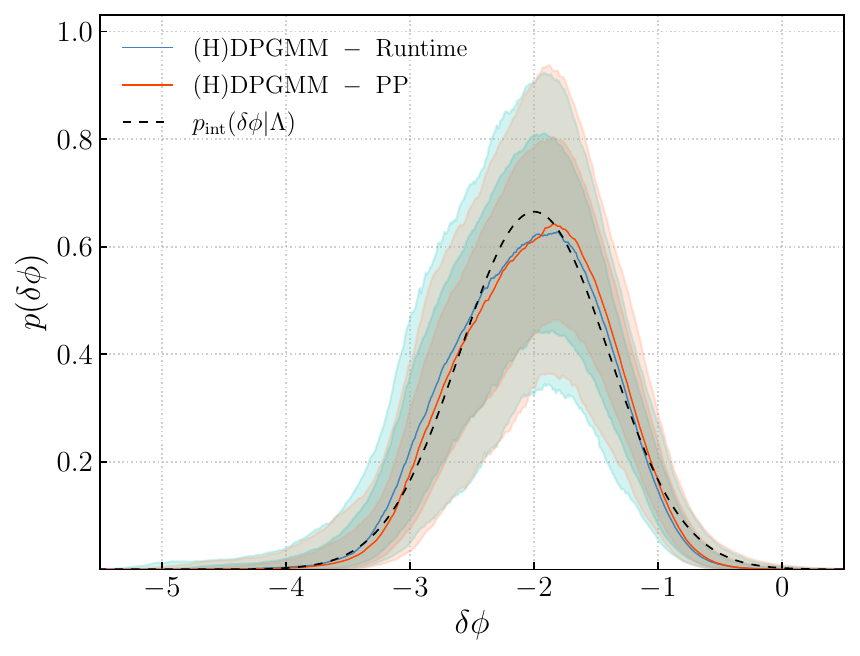}
    \caption{Non-parametric recovered posterior distribution for the intrinsic distribution (\emph{narrow} model, $\sigma_0=0.3$) accounting for the presence of selection biases using the two approaches presented in this paper: at runtime including the selection effects in the likelihood (orange) and in post-processing (blue). The black dashed line represents the true intrinsic distribution $\pint(\delta\phi|\Lambda)$.}
    \label{fig:selfunc_figaro}
\end{figure}

The conditioned probability distribution for the weights without accounting for selection effects is a Dirichlet distribution (DD):
\begin{equation}
p(\mathbf{w}|\mathbf{z}) = \mathrm{DD}(\mathbf{w}|\mathbf{N} + a/K)\,.
\end{equation}
Here $K$ is the number of Gaussian components, $\mathbf{N} = \{N_1,\ldots,N_K\}$ is the number of observations associated with each cluster and $a$ is the concentration parameter of the Dirichlet distribution. 
The presence of selection effects reduces the number of observations by a factor $\alpha(\mu,\sigma)$,
\begin{multline}
N_j =  N'_j \int \pdet(\theta)\mathcal{N}(\theta|\mu_j,\sigma_j)\dd \theta \equiv N'_j\alpha(\mu_j,\sigma_j)\\
\Rightarrow N'_j = \frac{N_j}{\alpha(\mu_j,\sigma_j)}\,,
\end{multline}
where $N_j'$ denotes the number of observations in absence of selection effects. Defining $\mathbf{N}' = \{N'_1,\ldots,N'_K\}$, the conditioned probability for $\mathbf{w}$ including selection effects becomes
\begin{equation}
p(\mathbf{w}|\mathbf{z},\boldsymbol\mu,\boldsymbol\sigma,\det) = \mathrm{DD}(\mathbf{w}|\mathbf{N}' + a/K)\,.
\end{equation}
In Figure~\ref{fig:selfunc_figaro} I show the non-parametric reconstruction of the \emph{narrow} population model with $\sigma_0 = 0.3$ with the in-likelihood approach (\emph{runtime}) and in post-processing (PP). The two distributions are both consistent with the intrinsic distribution and they overlap almost perfectly, in agreement with the equivalence of the two ways of accounting for selection effects presented in this paper.

\section{Detection probability as a function of true parameters}\label{app:selfunc}
Ref.~\citep{essick:2023}, in the examples given in Section~4.2 and Appendix~B, generates the mock catalog including events with a probability that depends on the observed parameters $\hat\theta_i$ only and not on the true value of the parameters $\theta_i$ as done in this paper. At the same time, the self-consistent procedure provided in Appendix~A of Ref.~\citep{essick:2023} selects the events to include using a probability which is a function of true parameters $\theta_i$, modelling a deterministic threshold-based process.
Here I briefly summarise the approach taken here, showing its consistency with the recipe provided in the aforementioned Appendix, as well as showing that modelling the detection probability as a function of the observed parameters is inconsistent with a deterministic threshold process.

The data $d_i$ observed in the GW detector is the superposition of the signal $h(\theta_i)$, which by definition depends on the true parameters of the event, and the noise $n_i$:
\begin{equation}
    d_i = n_i + h(\theta_i)\,.
\end{equation}
The inclusion of a specific event in a catalogue, once the noise realisation is fixed, is a deterministic process. If a detection statistic $\detstat$, which depends on the data $d_i$ hence on the signal (the true parameters of the event) and the specific noise realisation in a deterministic way, passes a certain threshold value $\detstat_\mathrm{th}$, the event is considered part of the catalogue:
\begin{equation}\label{eq:deterministic_detection}
\pdet(\theta_i,n_i) = 
\begin{cases}
1 \qif \detstat(\theta_i,n_i) > \detstat_\mathrm{th}\\
0 \qif \detstat(\theta_i,n_i) < \detstat_\mathrm{th}
\end{cases}.
\end{equation}
This deterministic process is turned into a stochastic process when the noise realisation is marginalised out:
\begin{equation}
    \pdet(\theta_i) = \int \pdet(\theta_i,n_i)p(n_i)\dd n_i\,.
\end{equation}
An event with parameters $\theta_i$, drawn from the intrinsic distribution, is therefore included in the catalogue with a probability equal to $\pdet(\theta_i)$, depending on both $\theta_i$ and a random noise realisation $n_i$, hence on the data $d_i$ as in the physical DAG of Ref.~\citep{essick:2023}.
The available selection function approximants \citep{wysocki:2019,veske:2021,lorenzo:2024} are calibrated to reproduce the results of the LVK sensitivity estimate injection campaign \citep{sensitivityestimate:2021}.
The same variability is encoded in the prescription summarised in Appendix~A of Ref.~\citep{essick:2023} (step 1): a sample $\theta_i$ is drawn from the intrinsic distribution (step 1a) and then a detection statistic, the signal-to-noise ratio $\rho_\mathrm{obs}$, is drawn from a distribution that depends on $\theta_i$, the true parameters of the specific event (step 1b), \emph{de facto} drawing a noise realisation:
\begin{equation}
    \rho_\mathrm{obs,i} \sim p\qty(\rho_\mathrm{obs}|\rho_\mathrm{opt}(\theta_i))\,,
\end{equation}
The sample is kept if $\rho_\mathrm{obs,i} \geq \detstat_\mathrm{th}$ (step 1c). Alternatively, one can write
\begin{equation}
    \pdet(\theta_i) = \int_{\detstat_\mathrm{th}}^{\infty} p\qty(\rho_\mathrm{obs}|\rho_\mathrm{opt}(\theta_i)) \dd \rho_\mathrm{obs}\,.
\end{equation}

The \emph{observed} parameters for an individual event are drawn from a distribution that depends both on the true parameters of the events and the specific noise realisation associated with the event:
\begin{equation}
p(\hat\theta|\theta_i, n_i) = p(\hat\theta|d_i)\,.
\end{equation}
In other words, the observed parameters distribution is conditioned on the observed data $d_i$. The detection process, however, is deterministic once the noise realisation is fixed -- Eq.~\eqref{eq:deterministic_detection} of this paper and Eq.~(A1) of Ref.~\citep{essick:2023}. 
Selecting the observed events with a selection function that depends on the observed parameters $\hat\theta_i$ rather than on the true parameters $\theta_i$, as done in the example presented in Section~4.2 and Appendix~B of Ref.~\citep{essick:2023}, is not consistent with a deterministic threshold-based selection process: for example, considering the maximum-likelihood estimate for the parameters as $\hat\theta_i$ -- which is fixed once we consider the specific noise realisation $n_i$, hence the data $d_i$ -- and modelling the detection probability as some function $f$ of the observed parameters, 
\begin{equation}
\pdet(\hat\theta_i) = f(\hat\theta_i)\,,
\end{equation}
leads to the conclusion that the very same stretch of data $d_i$ might be flagged alternatively \emph{detected} and \emph{not detected} by two subsequents runs of a detection process. This behaviour is not expected in a deterministic threshold-based selection process.

Whereas modelling the detection procedure as a non-deterministic function of the detection statistic is not wrong \emph{per se} and the inclusion of such procedure in the likelihood function allows to correctly infer the population parameters $\Lambda$ as shown in Ref.~\citep{essick:2023}, examples built around this approach are not suitable as test bed for the framework presented in this paper.

\bibliography{bibliography.bib}
\end{document}